CONDENSED
MATTER

# Quantum Effects at a Spin-Flop Transition in the Antiferromagnetic Topological Insulator MnBi$_2$Te$_4$

V. V. Val'kov[a,*], A. O. Zlotnikov[a], A. Gamov[a], N. A. Fedorova[a], and F. N. Tomilin[a]

[a] *Kirensky Institute of Physics, Federal Research Center KSC, Siberian Branch, Russian Academy of Sciences, Krasnoyarsk, 660036 Russia*

*\*e-mail: vvv@iph.krasn.ru*

It is shown that the experimentally detected features in the low-temperature behavior of the magnetization in an external magnetic field perpendicular to the layers of manganese ions of the topological antiferromagnet MnBi$_2$Te$_4$ are due to quantum effects induced by the off-diagonal nature of the trigonal component of the crystal field. In this case, the anomalous increase in the magnetization of the material before the spin-flop transition, as well as after it in the phase of "collapsed" sublattices, is explained by the suppression of contributions from quantum effects. The comparison of the results of the theoretical analysis with experimental data has made it possible to refine the parameters of the effective spin model of MnBi$_2$Te$_4$ and to establish the important role of the noted trigonal component.

## 1. INTRODUCTION

The layered van der Waals material MnBi$_2$Te$_4$ is of great interest because the antiferromagnetic (AFM) order existing in this topological insulator below the Néel temperature $T_N = 24.5$ K [1–5] allows one to affect topologically protected surface states through the magnetic structure.

Magnetic ordering in MnBi$_2$Te$_4$ corresponds to type-A antiferromagnetism, where the magnetic moments of Mn$^{2+}$ ions within one layer form a ferromagnetic (FM) structure, while these moments in adjacent planes are antiparallel. An important feature of the magnetic properties of MnBi$_2$Te$_4$ is due to the strong single-ion anisotropy (SIA), according to which the magnetic moments of Mn ions are perpendicular to the layers. These facts are confirmed by neutron scattering data [6, 7]. At the same time, the magnetic and spectral properties of MnBi$_2$Te$_4$ may differ for different samples [8]. For example, the formation of ferromagnetic order was detected in some samples on the MnBi$_2$Te$_4$ surface at $T < T_N$ [9, 10].

The mechanism of the formation of ferromagnetism in Mn layers, which is not explained by the direct Heisenberg exchange, is of interest. Various estimates according to the Goodenough–Kanamori–Anderson rule suggest the formation of either weak AFM exchange or FM exchange between Mn ions in one layer through Te$^{2-}$ ligands [11, 12]. Ab initio calculations of the exchange parameters in MnBi$_2$Te$_4$ confirm the implementation of the FM exchange of 0.09 meV between the nearest Mn ions in the layer [2]. However, there is no commonly accepted reason for ferromagnetism and studies in this direction continue. It was shown that hybridization of $p$ states of Bi$^{3+}$ ions with $p$ states of the Te$^{2-}$ ligand can initiate the FM superexchange between Mn ions [12]. It was also noted that the ferromagnetic coupling between the magnetic moments of Mn ions located in one layer can be due to the kinematic interaction between Hubbard fermions, reflecting the behavior of a strongly correlated subsystem of Mn$^{2+}$3$d$ electrons [13]. In this case, the AFM exchange within the layer and between layers of Mn$^{2+}$ ions occurs via the Anderson mechanism.

The occurrence of the SIA in MnBi$_2$Te$_4$ remains problematic, since the orbital angular momentum in the 3$d^5$ electron configuration of the manganese ion and $S = 5/2$ is $L = 0$. One of the mechanisms for inducing the SIA may be the virtual mixing of other electron configurations. The authors of [11] showed that the spin–orbit coupling in Te ions is important for the onset of the SIA.

The MnBi$_2$Te$_4$ compound has the properties of an antiferromagnet with an "easy axis" SIA perpendicular to the manganese ion layers (the $c$ axis). This anisotropy stabilizes the collinear AFM structure (hereinafter phase I) in a magnetic field **H**, directed along the $c$ axis, up to the spin-flop transition field $H_{sf} = 3.7$ T, when the flop of magnetic sublattices occurs [2, 6, 14, 15] with the formation of a canted phase II. In this phase, an increase in $H$ leads to a decrease in the inclination angle of the magnetization vectors of the sublattices with respect to the $c$ axis. When $H$ reaches the saturation field $H_{\text{sat}}^{c} = 8.1$ T, the magnetizations of both sublattices are oriented along the $c$ axis (the phase III).

In the magnetic field lying in the plane of manganese ions, the sublattice canting occurs immediately, and their "collapse" occurs at the saturation field $H_{\text{sat}}^{ab} = 10.9$ T.

A phenomenological approach was used to describe the evolution of the magnetic structure and to estimate the exchange parameters from the characteristic fields [7, 11, 15]. However, some data did not agree with the phenomenological results. This primarily concerned the behavior of the magnetization $M$ at $T \ll T_N$, where the magnetization $M$ in the phase I increased with the magnetic field $H$, whereas the phenomenological theory predicts $M = 0$. A similar problem arose for a para process observed in the phase III.

The authors of [15] attributed the latter problem to defects in the system, when Mn ions occupy the positions of Bi ions (antisite defects). In this case, the AFM coupling is established between the Mn ions in the central layer of MnBi$_2$Te$_4$ and the "defective" Mn ions; as a result, a decrease in the temperature in the AFM phase leads to the establishment of a long-range order in the defective layer of Mn ions [15–17]. Indeed, the inclusion of the noted defects allowed one to explain the increase in the magnetization in the phase III with an increase in the field up to 60 T [15]. However, the issue of the increase in the magnetization in fields lower than the spin-flop transition field remained open.

It should be noted that the mixing of the positions of the Mn and Sb ions in the related MnSb$_2$Te$_4$ compound can lead to a change in the magnetic structure from type-A AFM to FM [16, 18]. In structures with MnBi$_2$Se$_4$, in addition to the defective MnI ions in the Bi ion layer, MnII ions were also found in the van der Waals gap between septuple layers [19]. At the same time, the exchange interaction of these ions with the Mn ions in the central layer differs: FM for MnI and AFM for MnII.

The solution of the above issues is the subject of this work with the focus on quantum effects arising in anisotropic magnets with relatively low magnetic ordering temperatures.

Since the internal fields for the crystallographic structure of MnBi$_2$Te$_4$ correspond to trigonal symmetry with the point group $C_{3d}$ ($S_6$) for Mn ions, the SIA operator contains a trigonal component in addition to the usual uniaxial terms [20, 21]. Its feature is due to the mixing of states with different spin projections. This causes quantum effects that are absent in the phenomenological approach.

The aim of the work is to develop a theory to describe the above anomalies in the behavior of the low-temperature magnetization considering quantum effects. To this end, we use an atomic representation [22] involving the diagram technique for Hubbard operators [23, 24], which significantly differs from the standard technique for Fermi or Bose operators [25] because the commutator of Hubbard operators may be another Hubbard operator rather than a number. This makes it possible to derive the dispersion relation and obtain its solution in the low-temperature region taking into account the renormalizations from the trigonal component of the SIA (TCSIA). Critical fields are determined and the behavior of the material magnetization is analyzed in all phases at different $T$ values. The comparison of theoretical dependences $M(H)$ with experimental data allows us to obtain additional information on the model parameters of the MnBi$_2$Te$_4$ magnetic subsystem and to demonstrate a significant role of the TCSIA.

## 2. HAMILTONIAN OF THE MAGNETIC SUBSYSTEM

Taking into account the presentation in the Introduction, we write the Hamiltonian of the magnetic subsystem of MnBi$_2$Te$_4$ in the form

$$\hat{H} = -\frac{1}{2}\sum_{ff'}I_{ff'}(\mathbf{S}_f\mathbf{S}_{f'}) - \frac{1}{2}\sum_{gg'}I_{gg'}(\mathbf{S}_g\mathbf{S}_{g'}) + \sum_{fg}J_{fg}(\mathbf{S}_f\mathbf{S}_g) + \sum_f[H_f^A - \mathbf{H}\mathbf{S}_f] + \sum_g[H_g^A - \mathbf{H}\mathbf{S}_g]. \quad (1)$$

Here, the subscripts $f$ and $f'$ enumerate the sites of the $F$ sublattice where the spins are oriented along the $Oz$ axis in zero magnetic field; the subscripts $g$ and $g'$ enumerate the sites of the $G$ sublattice, where the orientation of the spins is opposite; the parameters $I_{ff'}$ and $I_{gg'}$ are the exchange coupling constants between spins from the same sublattice located within the same layer; $J_{fg}$ is the exchange integral for spins from different sublattices; $\mathbf{S}_f$ and $\mathbf{S}_g$ are the vector spin operators at sites $f$ and $g$, respectively; and $\mathbf{H}$ is the magnetic field vector, whose magnitude is measured in energy units; i.e., $H$ means below the product $g_L\mu_B H$, where $g_L$ is the Landé factor and $\mu_B$ is the Bohr magneton.

In accordance with the crystal symmetry, the SIA energy operator for the $F$ sublattice given by the expression

$$H_f^A = -D_2(\hat{S}_f^z)^2 - D_4(\hat{S}_f^z)^4 + B\hat{\Omega}_{4f}^3 \quad (2)$$

includes the TCSIA with the constant $B$ and $\Omega_4^3$ (the site subscript $f$ is temporarily omitted) has the form [20]

$$\hat{\Omega}_4^3 = (\hat{O}_4^{+3} - \hat{O}_4^{-3})/2i, \quad \hat{O}_4^{\pm 3} = (\hat{S}^z \hat{S}^{\pm 3} + \hat{S}^{\pm 3} \hat{S}^z)/2,$$
$$\hat{S}^{\pm 3} = (\hat{S}^{\pm})^3, \quad \hat{S}^{\pm} = \hat{S}^x \pm i\hat{S}^y, \quad (3)$$

where $\hat{S}^x$, $\hat{S}^y$, and $\hat{S}^z$ are the $x$, $y$, and $z$ components of the spin operator vector. The operator $H_g^A$ is given by Eq. (2), where the subscript $f$ is replaced by the subscript $g$.

Phenomenologically, the positive parameters $D_2$ and $D_4$ define the intensities of effective fields that orient the spins either along or against the $z$ direction. At the negative parameters $D_2$ and $D_4$, the effective fields tend to order the spins perpendicular to the $Oz$ axis. The constant $B$ sets the intensity of the effective anisotropy field corresponding to the second axis (a biaxial crystal).

In the quantum approach, the SIA parameters correspond to the amplitudes of the quadrupole and octupole fields that act on the spins and arise due to the combined effect of the crystal field and the spin–orbit coupling.

### 3. ATOMIC REPRESENTATION, SINGLE-SITE STATES, AND SELF-CONSISTENT EQUATIONS

Since the magnetic ordering temperature in $MnBi_2Te_4$ is low ($T_N \simeq 24$ K), the characteristic SIA energy can become comparable with the exchange interaction energy. Under these conditions, the dynamics of the magnetic subsystem is determined not only by the dipole moments, but also by the dynamics of higher multipoles. To describe such systems, an extended basis is used, which includes the degrees of freedom associated with spin operators and their admissible products.

In its most general form, such a program is implemented in the atomic representation, where the operators of the group $U(N)$ become the dynamic variables. To this end, the states $|\Psi_f^F\rangle$ and $|\Psi_g^G\rangle$ are introduced as solutions of the Schrödinger equations

$$(H_f^A - \bar{\mathbf{H}}_F \mathbf{S}_f)|\Psi_{nf}^F\rangle = E_n^F |\Psi_{nf}^F\rangle, \quad n = 1, 2, \ldots, 6, \quad (4)$$

$$(H_g^A - \bar{\mathbf{H}}_G \mathbf{S}_g)|\Psi_{ng}^G\rangle = E_n^G |\Psi_{ng}^G\rangle, \quad n = 1, 2, \ldots, 6, \quad (5)$$

where

$$\bar{\mathbf{H}}_F = \mathbf{H} + I_0 \boldsymbol{\sigma}_F - J_0 \boldsymbol{\sigma}_G,$$
$$\bar{\mathbf{H}}_G = \mathbf{H} + I_0 \boldsymbol{\sigma}_G - J_0 \boldsymbol{\sigma}_F \quad (6)$$

are the self-consistent field vectors depending on equilibrium averages $\boldsymbol{\sigma}_F = \langle \mathbf{S}_f \rangle$ and $\boldsymbol{\sigma}_G = \langle \mathbf{S}_g \rangle$ for the $F$ and $G$ sublattices, respectively. In Eq. (6), $I_0$ and $J_0$ are the Fourier transforms of the exchange integrals $I_{ff'}$ and $J_{fg}$ at the zero quasimomentum, respectively.

The equilibrium averages satisfy the equations

$$\boldsymbol{\sigma}_F = \sum_{n=1}^{6} \langle \Psi_n^F | \mathbf{S}_f | \Psi_n^F \rangle \exp(-E_n^F/T)/Z_F,$$

$$\boldsymbol{\sigma}_G = \sum_{n=1}^{6} \langle \Psi_n^G | \mathbf{S}_g | \Psi_n^G \rangle \exp(-E_n^G/T)/Z_G, \quad (7)$$

$$Z_F = \sum_{n=1}^{6} \exp(-E_n^F/T), \quad Z_G = \sum_{n=1}^{6} \exp(-E_n^G/T).$$

Expressions (4)–(7) form a closed system of self-consistent equations that determine the necessary thermodynamic averages.

If the single-site states $|\Psi_{nf}^F\rangle$ and $|\Psi_{ng}^G\rangle$ are considered as the basis vectors of the Hilbert space, the transition to the atomic representation is carried out by introducing the Hubbard operators [22]

$$X_f^{np} = |\Psi_{nf}^F\rangle\langle\Psi_{pf}^F|, \quad X_g^{mq} = |\Psi_{mg}^G\rangle\langle\Psi_{qg}^G|, \quad (8)$$

which serve as new dynamic variables. In terms of these variables, the dynamics of not only dipole but also higher degrees of freedom is described. The relation of spin operators with new variables is written as

$$S_f^+ = \sum_{np} \gamma_{np}^F X_f^{np} \to \sum_{\alpha} \gamma_{\alpha}^F X_f^{\alpha}, \quad (9)$$

where for brevity, a pair of subscripts in the second sum is denoted by $\alpha$. The representation parameters are calculated as the matrix elements

$$\gamma_{np}^F = \langle \Psi_{nf}^F | S_f^+ | \Psi_{pf}^F \rangle. \quad (10)$$

The other operators are written in a similar form

$$S_f^z = \sum_{\mu} \Gamma_{\mu}^F X_f^{\mu}, \quad S_g^+ = \sum_{\beta} \gamma_{\beta}^G X_g^{\beta},$$
$$S_g^z = \sum_{\nu} \Gamma_{\nu}^G X_g^{\nu}. \quad (11)$$

In the new variables, the Hamiltonian takes the form

$$\hat{H} = \sum_{fn} E_n^F X_f^{nn} + \sum_{gm} E_m^G X_g^{mm} + H_{\text{int}}, \quad (12)$$

where $H_{\text{int}}$ is obtained from the exchange interaction operators into which Eqs. (9) and (11) for the spin operators should be substituted. The Hamiltonian specified by Eq. (12) allows one to use the diagram technique for Hubbard operators [23, 24].

## 4. SINGLE-SITE STATES AND ENERGIES IN THE COLLINEAR GEOMETRY

In the magnetic field directed along the trigonal axis in the phase I, the single-site states for the $F$ sublattice have the form (the site subscript $f$ is omitted)

$$|\Psi_1^F\rangle = \cos\alpha_+|5/2\rangle + i\sin\alpha_+|-1/2\rangle,$$

$$|\Psi_2^F\rangle = |3/2\rangle, \quad |\Psi_5^F\rangle = |-3/2\rangle,$$

$$|\Psi_3^F\rangle = \cos\alpha_-|1/2\rangle + i\sin\alpha_-|-5/2\rangle, \quad (13)$$

$$|\Psi_4^F\rangle = i\sin\alpha_+|5/2\rangle + \cos\alpha_+|-1/2\rangle,$$

$$|\Psi_6^F\rangle = i\sin\alpha_-|1/2\rangle + \cos\alpha_-|-5/2\rangle,$$

where $|m\rangle$, $(m = 5/2, 3/2, ..., -5/2)$ are the eigenstates of the operator $\hat{S}^z$ for the spin $S = 5/2$,

$$\cos\alpha_\pm = \sqrt{\frac{1+\delta_\pm}{2}}, \quad \sin\alpha_\pm = \mp\frac{V}{|V|}\sqrt{\frac{1-\delta_\pm}{2}},$$

$$\delta_\pm = \frac{\Delta_\pm}{\sqrt{\Delta_\pm^2 + V^2}}, \quad \Delta_\pm = [3\bar{H}_F \pm (6D_2 + 39D_4)]/2,$$

$$\bar{H}_F = H + I_0\sigma_F - J_0\sigma_G, \quad V = 3\sqrt{10}B. \quad (14)$$

The following eigenenergies correspond to the single-site states given by Eqs. (13):

$$E_{1,4}^F = (\varepsilon_4^F + \varepsilon_1^F)/2 \mp \sqrt{\Delta_+^2 + V^2}, \quad E_2^F = \varepsilon_2^F,$$

$$E_{3,6}^F = (\varepsilon_6^F + \varepsilon_3^F)/2 \mp \sqrt{\Delta_-^2 + V^2}, \quad E_5^F = \varepsilon_5^F, \quad (15)$$

$$\varepsilon_n^F = -\bar{H}_F(7/2 - n) - D_2(7/2 - n)^2 - D_4(7/2 - n)^4.$$

For the $G$ sublattice, the single-site states are similar:

$$|\Psi_1^G\rangle = \cos\beta_+|-5/2\rangle - i\sin\beta_+|+1/2\rangle,$$

$$|\Psi_2^G\rangle = |-3/2\rangle, \quad |\Psi_5^G\rangle = |+3/2\rangle,$$

$$|\Psi_3^G\rangle = \cos\beta_-|-1/2\rangle + i\sin\beta_-|+5/2\rangle, \quad (16)$$

$$|\Psi_4^G\rangle = -i\sin\beta_+|-5/2\rangle + \cos\beta_+|+1/2\rangle,$$

$$|\Psi_6^G\rangle = i\sin\beta_-|-1/2\rangle + \cos\beta_-|+5/2\rangle,$$

where

$$\cos\beta_\pm = \sqrt{\frac{1+\delta_\pm^G}{2}}, \quad \sin\beta_\pm = \frac{V}{|V|}\sqrt{\frac{1-\delta_\pm^G}{2}},$$

$$\delta_\pm^G = \frac{\Delta_\pm^G}{\sqrt{(\Delta_\pm^G)^2 + V^2}}, \quad \Delta_\pm^G = \frac{3\bar{H}_G \pm (6D_2 + 39D_4)}{2},$$

$$\bar{H}_G = -H - I_0\sigma_G + J_0\sigma_F. \quad (17)$$

Single-site energies for the $G$ sublattice are specified by the expressions

$$E_{1,4}^G = (\varepsilon_4^G + \varepsilon_1^G)/2 \mp \sqrt{(\Delta_+^G)^2 + V^2}, \quad E_2^G = \varepsilon_2^G,$$

$$E_{3,6}^G = (\varepsilon_6^G + \varepsilon_3^G)/2 \mp \sqrt{(\Delta_-^G)^2 + V^2}, \quad E_5^G = \varepsilon_5^G, \quad (18)$$

$$\varepsilon_m^G = \bar{H}_G(m - 7/2) - D_2(m - 7/2)^2 - D_4(m - 7/2)^4.$$

## 5. DISPERSION RELATION

The increase in the field $H$ applied along the easy axis of anisotropy induces a first-order phase transition [26, 27] from the phase I to the phase II. The spin-flop transition field $H_{sf}$ is found from the condition of loss of positive definiteness of the excitation spectrum $\Omega_k$ in the phase I.

The spin-flop transition field $H_{sf}$ calculated using the phenomenological approach may not coincide with the quantum theory value, in particular, because of quantum fluctuations that are manifested at low temperatures and especially in low-dimensional magnets [28], as well as in compounds with a triangular lattice. In this case, frustrations increase the effect of quantum corrections [29, 30], which can lead to additional features in the behavior of the magnetization.

Taking into account the said above, when calculating $\Omega_k$ under strong SIA conditions, we apply the Matsubara Green's functions constructed in the Hubbard operator basis [23, 24]:

$$D_{\alpha\beta}^{AB}(l - l', \tau - \tau') = -\langle\hat{T}_\tau\tilde{X}_l^\alpha(\tau)\tilde{X}_{l'}^{-\beta}(\tau')\rangle$$

$$= \frac{T}{N}\sum_{k\omega_m}\exp[ik(l - l') - i\omega_m(\tau - \tau')]D_{\alpha\beta}^{AB}(k, \omega_m). \quad (19)$$

Here, the angle brackets denote statistical averaging of the product of Hubbard operators ordered in the Matsubara time (the effect of the operator $\hat{T}_\tau$) taken in the Heisenberg representation at the Matsubara times $\tau$ and $\tau'$. The sites $l$ and $l'$ refer to the $A$ and $B$ sublattices, respectively, $D_{\alpha\beta}^{AB}(k, \omega_m)$ is the Fourier transform of the Matsubara Green's functions, $k$ is the quasimomentum, and $\omega_m = 2m\pi T(m = 0, \pm 1, \pm 2, ...)$ is the Matsubara frequency [25].

The diagram technique for Hubbard operators method is described in [23, 24, 31, 32], therefore, in Fig. 1 we present only a graphical representation of the system of equations for $G_{\alpha\beta}^{AB}(k, \omega_m)$ in the approximation of noninteracting quasiparticles. Two lines with arrows denote desired functions $G_{\alpha\beta}^{AB}(k, \omega_m)$, which are related to $D_{\alpha\beta}^{AB}(k, \omega_m)$ as $D_{\alpha\beta}^{AB} = G_{\alpha\beta}^{AB}b(\beta)$, where $b(\beta)$ is the end factor [23, 24] shown by a circle. Lines with an

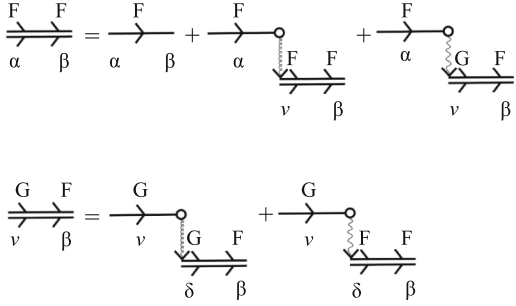

**Fig. 1.** System of equations for $G^{AB}_{\alpha\beta}(k,\omega_m)$.

arrow correspond to a single-site propagator for the $A$ sublattice:

$$G^A_\alpha(i\omega_m) = (i\omega_m + E^A_p - E^A_q)^{-1}, \quad \alpha \equiv \alpha(p,q). \quad (20)$$

The dotted and wavy lines denote the exchange integrals $I_k$ and $J_k$, respectively.

After comparing the graphs of analytical expressions, we conclude that

$$G^{FF}_{\alpha\beta}(p) = \delta_{\alpha\beta} G^F_\alpha(i\omega_m)$$
$$-G^F_\alpha(i\omega_m)b(\alpha)(\gamma^F_\alpha)^*[I_k X_\beta(p) - J_k Y_\beta(p)]/2, \quad (21)$$

$$G^{GF}_{\nu\beta}(p) = G^G_\nu(i\omega_m)b(\nu)(\gamma^G_\nu)^*[J_k X_\beta(p) - I_k Y_\beta(p)]/2,$$

where $p \equiv (k, i\omega_m)$ and

$$X_\beta(p) = \sum_\nu \gamma^F_\nu G^{FF}_{\nu\beta}(p), \quad Y_\beta(p) = \sum_\nu \gamma^G_\nu G^{GF}_{\nu\beta}(p).$$

Solving Eq. (21), we obtain

$$D^{FF}_{\alpha\alpha}(p) = D^F_\alpha(i\omega_m)[1 + R^F_\alpha(p)]. \quad (22)$$

Here, $(D^F_\alpha(i\omega_m) = G^F_\alpha(i\omega_m)b^F(\alpha))$,

$$R^F_\alpha(p) = \frac{|\gamma^F_\alpha|^2 D^F_\alpha \{J_k^2 \Phi^G/4 - I_k[1 + I_k \Phi^G/2]/2\}}{\det(p)}, \quad (23)$$

where $I_k$ and $J_k$ are the Fourier transforms of exchange integrals given by the expressions

$$I_k = 2I[\cos(k_1) + \cos(k_2) + \cos(k_1 + k_2)], \quad (24)$$

$$J_k = 2J[\cos(k_3/2) + \cos(k_3/2 + k_1) + \cos(k_3/2 + k_1 + k_2)]. \quad (25)$$

Here, $k_1, k_2$, and $k_3$ are the components of the quasi-momentum belonging to the first Brillouin zone. The basis vectors of the triangular lattice in the planes of Mn ions are denoted as $\mathbf{a}_1$ and $\mathbf{a}_2$, and the basis vector of the unit cell $\mathbf{a}_3$ is perpendicular to these planes. The

functions $\Phi^F$ and $\Phi^G$ in Eq. (23) are defined by the formulas

$$\Phi^F(i\omega_m) = \sum_{nl} \frac{|\gamma^F(nl)|^2 (N^F_n - N^F_l)}{i\omega_m + E^F_n - E^F_l}, \quad (26)$$

$$\Phi^G(i\omega_m) = \sum_{nl} \frac{|\gamma^G(nl)|^2 (N^G_n - N^G_l)}{i\omega_m + E^G_n - E^G_l}, \quad (27)$$

where $N^F_n$ and $N^G_n$ are the filling numbers of the single-site states for the $F$ and $G$ sublattices, respectively. The denominator in Eq. (23) has the form

$$\det(p) = [1 + I_k \Phi^F/2][1 + I_k \Phi^G/2] - J_k^2 \Phi^F \Phi^G/4. \quad (28)$$

Its analytical continuation gives the following equation for calculating the excitation spectrum:

$$\det(k, i\omega_m \to \omega + i\delta) = 0, \quad \delta \to +0. \quad (29)$$

## 6. QUANTUM EFFECTS AND AFM MAGNETIZATION IN PHASES I AND III AT $T \ll T_N$

It is known that the magnetization is zero up to exponentially small terms and does not depend on $H$ in uniaxial antiferromagnets at low temperatures in the phase I in a magnetic field directed along the anisotropy axis. The independence of the magnetization of $H$ holds with the same accuracy also in the phase III at $H > H^c_{\text{sat}}$ ($H^c_{\text{sat}}$ is the critical saturation field), when the sublattices are aligned along the easy axis.

The inclusion of the TCSIA qualitatively changes the situation. In this case, a change in $M$ is described by power-law rather than exponentially small terms. Indeed, at $T \ll T_N$, it follows from Eqs. (13) and (16) that the system of self-consistent equations for $\sigma_F$ and $\sigma_G$ has the form

$$\sigma_F = S - 3\sin^2\alpha_+, \quad \sigma_G = -S + 3\sin^2\beta_+. \quad (30)$$

Its solution for the phase I in approximations quadratic in the parameter $B$ and linear in $H$ and taking into account Eqs. (14) and (17) gives the following expression for the magnetization $M(H,T) = \sigma_F + \sigma_G$ (in units of $\mu_B$) per Mn ion:

$$M(H) = \left(\frac{192}{25}\right)\left(\frac{B}{I_+}\right)^2\left(\frac{H}{I_+}\right), \quad I_+ = I_0 + J_0. \quad (31)$$

It can be seen that the TCSIA, due to quantum effects, leads to an increase in the magnetization with increasing $H$.

In the phase III, $\sigma_F = \sigma_G$ and the magnetization has the form

$$M(H) = 2\left[\frac{5}{2} - \frac{24}{5}\left(\frac{B}{I_+}\right)^2 + \frac{96}{25}\left(\frac{B}{I_+}\right)^2\left(\frac{\delta H}{I_+}\right)\right], \quad (32)$$

where $\delta H = H - 2SJ_0$, the second term gives the decrease in the magnetization due to quantum effects, and the third term describes the para process caused by the suppression of the noted effects by the magnetic field. The physical reason for quantum renormalizations is the mixture of states with different spin projections by the operator $\hat{\Omega}_4^3$.

These results require an important remark. Recall that the Néel state corresponds to the eigenfunction of the Hamiltonian given by Eq. (1) only approximately even without the TCSIA. Therefore, the system contains so-called quantum zero-point fluctuations (QZPFs), which reduce the magnetization of sublattices at $T = 0$.

In uniaxial antiferromagnets, the magnetic field applied along the easy axis in the phase I acts differently on the spins of the $F$ and $G$ sublattices. For spins of the $F$ sublattice, the field $H$ favors ordering, while for the $G$ sublattice, it acts in the opposite way. Therefore, one would expect that QZPFs at $H \neq 0$ can also make an additional contribution to the magnetic-field dependence of the magnetization.

To clarify this problem, we calculate the contribution of QZPFs to the magnetization of a uniaxial (for simplicity) antiferromagnet in a longitudinal magnetic field in the phase I.

Using the Green's function given Eq. (22) and the well-known procedure to calculate averages, after summation over the Matsubara frequencies [25], we obtain the following expression for the magnetization of the $F$ sublattice per site:

$$\sigma_F = S - \frac{1}{2N}\sum_k \frac{A_k - \omega_k}{\omega_k}$$
$$-\frac{1}{2N}\sum_k\left\{\frac{A_k - \omega_k}{\omega_k}f\left(\frac{\Omega_k^-}{T}\right) + \frac{A_k + \omega_k}{\omega_k}f\left(\frac{\Omega_k^+}{T}\right)\right\}, \quad (33)$$

where $f(x)$ is the Bose–Einstein distribution function,

$$\Omega_k^{\pm} = \omega_k \pm H, \quad \omega_k = \sqrt{[A_k - SJ_k][A_k + SJ_k]},$$
$$A_k = \varepsilon_a + S(J_0 + I_0 - I_k), \quad \varepsilon_a = 4D_2 + 34D_4. \quad (34)$$

For the magnetization of the $G$ sublattice, we obtain

$$\sigma_G = -S + \frac{1}{2N}\sum_k \frac{A_k - \omega_k}{\omega_k}$$
$$+\frac{1}{2N}\sum_k\left\{\frac{A_k + \omega_k}{\omega_k}f\left(\frac{\Omega_k^-}{T}\right) + \frac{A_k - \omega_k}{\omega_k}f\left(\frac{\Omega_k^+}{T}\right)\right\}. \quad (35)$$

The second terms of Eqs. (33) and (35) are due to QZPFs, and the third terms describe the decrease in the magnetizations of sublattices due to thermal excitations of quasiparticles.

It is important that the contribution from QZPFs does not depend on $H$. As a result, the magnetization of a magnet per Mn ion at $T \ll T_N$ is independent of QZPFs and is determined only by thermal fluctuations:

$$M(H,T) = \frac{1}{N}\sum_k\left\{f\left(\frac{\Omega_k^-}{T}\right) - f\left(\frac{\Omega_k^+}{T}\right)\right\}. \quad (36)$$

However, this formula gives an increase in $M$ with the increase in the field $H$ in the range of $0 < H < H_{sf}$, significantly less than that observed in the experiment [15].

At $B \neq 0$, QZPFs begin to contribute to the magnetization, but this contribution is much smaller than the contribution from Eq. (31) at real SIA parameters because, in addition to smallness over $B$ and $H$, QZPFs introduce additional smallness due to the expansion in $1/r_0^3$.

Thus, the problem of the contribution from QZPFs to the dependence $M(H)$ in the phase I is solved.

## 7. RENORMALIZATION OF CRITICAL FIELDS

The boundary of the phase I implementation region is determined by the spin-flop transition field $H_{sf}$, which is evaluated from the condition that the excitation energy vanishes at $k = 0$. From Eqs. (26) and (27) at $T \ll T_N$, we obtain

$$\Phi^F(i\omega_m) = \frac{2\tilde{S}}{i\omega_m - E_{21}^F}, \quad \Phi^G(i\omega_m) = \frac{-2\tilde{S}}{i\omega_m + E_{21}^G}, \quad (37)$$

where

$$\tilde{S} = S(1+\xi), \quad \xi = (12/25)(B/I_+)^2,$$
$$E_{21}^F = H + \varepsilon_a + SI_+ + 36B^2/(5I_+), \quad (38)$$
$$E_{21}^G = -H + \varepsilon_a + SI_+ + 36B^2/(5I_+).$$

The renormalization $\sim\xi$ is due to the interplay between the spin dynamics and the dynamics of higher multipoles. Entanglement of these degrees of freedom occurs because of the TCSIA. In this case, the dispersion relation takes the form

$$[\omega - E_{21}^F + \tilde{S}I_k][\omega + E_{21}^G - \tilde{S}I_k] + (\tilde{S}J_k)^2 = 0. \quad (39)$$

Thus, we obtain two branches of the spectrum in the phase I:

$$\Omega_{1k} = \tilde{\omega}_k - H, \quad \Omega_{2k} = \tilde{\omega}_k + H, \quad (40)$$

$$\tilde{\omega}_k = \sqrt{[\tilde{\varepsilon}_a + E_k^-][\tilde{\varepsilon}_a + E_k^+]}, \quad (41)$$

$$E_k^{\mp} = \tilde{S}(I_0 - I_k + J_0 \mp J_k), \quad \tilde{\varepsilon}_a = \varepsilon_a + 6B^2/I_+. \quad (42)$$

From this, the expression for the critical field follows

$$H_{sf} = \sqrt{\tilde{\varepsilon}_a(\tilde{\varepsilon}_a + 2\tilde{S}J_0)}. \quad (43)$$

In the phase III, the sublattices are identical, so

$$\Phi^F(i\omega_m) = \Phi^G(i\omega_m) = 2\tilde{S}/(i\omega_m - E_{21}), \quad (44)$$

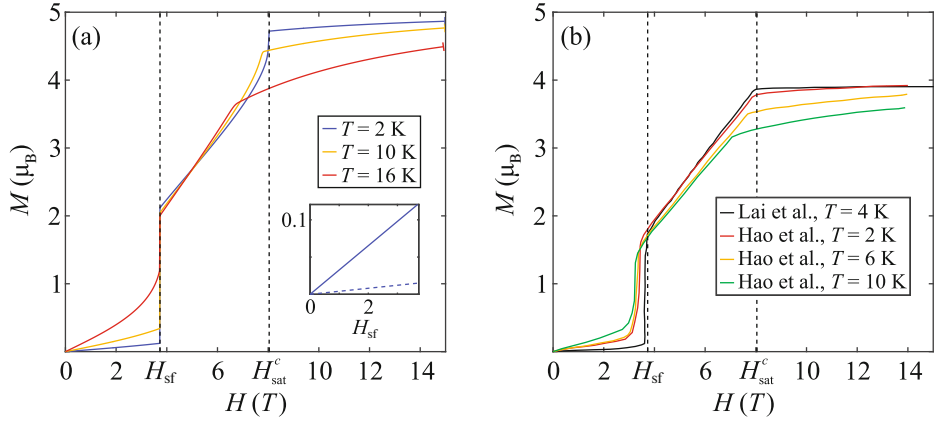

**Fig. 2.** (Color online) (a) Theoretically obtained $z$ component of the total magnetization of sublattices $M^z$ versus the magnetic field $H$, applied along the $Oz$ axis at temperatures $T = 2$, 10, and 16 K as calculated with the parameters $I_0 = 0.516$ meV, $J_0 = 0.219$ meV, $D_2 = 0.0095$ meV, $D_4 = 0$, and $B = 0.12$ meV (see the main text). The solid line in the inset shows the dependence $M^z(H)$ for $H \leq H_{sf}$, calculated by Eq. (31). The dashed line shows the dependence without the trigonal component of the crystal field ($B = 0$), see Eq. (36). (b) Experimental data for field dependences of the magnetization in the field $\mathbf{H} \parallel c$ for MnBi$_2$Te$_4$ from [14, 15] at temperatures $T = 2$, 4, 6, and 10 K.

and two branches of the spectrum are determined by the expressions

$$\Omega_k^\pm = H - H_{sat}^c + \tilde{S}[I_0 - I_k + J_0 \pm J_k], \quad (45)$$

where the second critical field has the form

$$H_{sat}^c = 2\sigma J_0 - \tilde{\varepsilon}_a, \quad \sigma = S - (24/5)(B/I_+)^2. \quad (46)$$

It is evident that renormalizations narrow the region of the phase II implementation.

## 8. FEATURES OF $M(H)$ AT THE SPIN-FLOP TRANSITION

When solving equations for $\sigma_F$ and $\sigma_G$, the model parameters were chosen taking into account four conditions. Two of them are determined by Eqs. (43) and (46) for critical fields. The resulting values had to coincide with the experimental data $H_{sf} = 3.7$ T and $H_{sat}^c = 8.1$ T [14, 15].

The third constraint relates the sum of the exchange integrals $I_0$ and $J_0$ to the mean-field Néel temperature: $T_N = S(S+1)(I_0 + J_0)/3$. Finally, the parameter $B$ was determined such that the para process occurs in the phase III at $T \ll T_N$ and the magnetization in the phase I increases noticeably, but the phase II remains.

Figure 2a shows the results of solving self-consistent Eqs. (4)–(7) demonstrating the important role of quantum effects in interpreting the features of the magnetization behavior detected in MnBi$_2$Te$_4$.

In the phases I and III, $M$ increases with the magnetic field $H$ at $T \ll T_N$. This is consistent with the experimental data shown in Fig. 2b. As noted above, such a dependence is physically due to quantum effects initiated by the TCSIA. At increasing temperature, contributions from thermal fluctuations are added to the noted effect. This explains the stronger dependence $M(H)$ at high temperatures.

Figure 3 shows fragments of the positions of the Mn ions forming the three nearest layers. In each of them, the ions are located at the site of triangular lattices. The left panel of Fig. 3 corresponds to the phase I, where the spins of the $F$ sublattice are in the middle layer (red solid circles) and are oriented along the $Oz$ axis. The projections of the positions of the Mn ions from the lower and upper layers on the middle layer are shown by blue and empty circles, respectively. The spins of these ions are oriented against the $Oz$ axis ($G$ sublattice). The middle panel shows the spin configuration in the phase II, where the $z$ projections of spins of both sublattices are oriented along the $Oz$ axis, and the transverse components for $F$ and $G$ sublattices are denoted by red and blue arrows, respectively. In the phase III (right panel), all spins are oriented along the $Oz$ axis.

The comparison of the dependences $M(H)$ shown in Figs. 2a and 2b shows that the theoretical magnetization in the phase III is close to the nominal one, but is larger than the experimental one. Such a discrepancy was previously explained by the influence of defects [15], which are disregarded in the model under consideration. In this regard, we note the work [19], where important results on the nature of disordering and the influence of non-ideality on magnetic properties were obtained for an AFM topological insulator heterostructure prepared on the basis of MnBi$_2$Se$_4$ and $Bi_2Te_3$. In particular, it was found that manganese ions can occupy positions different from the nominal ones. In this case, magnetic interactions between Mn and Se ions arise and induce a magnetic moment on Se ions. These data indicate the need to generalize the considered model in order to take into account defects, near which the spin density is redistributed. As a result, one could expect a decrease in $M$. However, the consideration of this important issue is beyond the scope of this work.

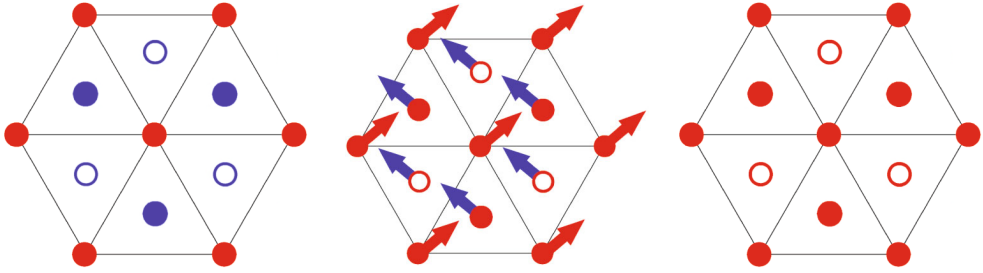

**Fig. 3.** (Color online) Magnetic structure in the phases (from left to right) I, II, and III. Red and blue colors correspond to the positive and negative components of the magnetization along the $Oz$ axis, respectively. The arrows in the middle panel define the magnetization vector in the $M_\perp = (M^x, M^y)$ plane.

A new feature caused by the TCSIA is manifested in the phase II (see Fig. 4): in addition to $\sigma_F^x$ and $\sigma_G^x = -\sigma_F^x$, $\sigma_F^y$ and $\sigma_G^y = \sigma_F^y$ are nonzero (the $Ox$ axis is parallel to the basis vector $\mathbf{a}_1$ of the triangular lattice). Phenomenologically, the emerging geometry of the magnetization vectors of the sublattices is explained by the biaxiality of the crystal, at which an "effective" anisotropy field appears, the direction of which does not coincide with the direction of the vector $\mathbf{a}_1$.

The field dependences of nonzero components of magnetization sublattices, when passing through three phases, are shown in Fig. 4.

## 9. CONCLUSIONS

**1.** It has been shown that quantum effects initiated by the trigonal component of the single-ion anisotropy lead to the emergence of magnetization dependences on the magnetic field in the low-temperature region both before the spin-flop transition and in the phase of collapsed sublattices. At the same time, the para process in the phase III, where the magnetizations of the sublattices are oriented along the $Oz$ axis, occurs because the magnetic field suppresses quantum effects and leads to an experimentally observed increase in the magnetization. This corresponds to the features in the dependences $M(H)$, discovered when studying the magnetic properties of the antiferromagnetic topological insulator $MnBi_2Te_4$.

**2.** The theoretical results obtained using the atomic representation and the diagram technique for Hubbard operators have demonstrated an important role of quantum effects. This approach has allowed us to correctly describe the influence of the strong single-ion anisotropy and to obtain a dispersion relation that determines the excitation spectrum taking into account the off-diagonal nature of the single-ion anisotropy.

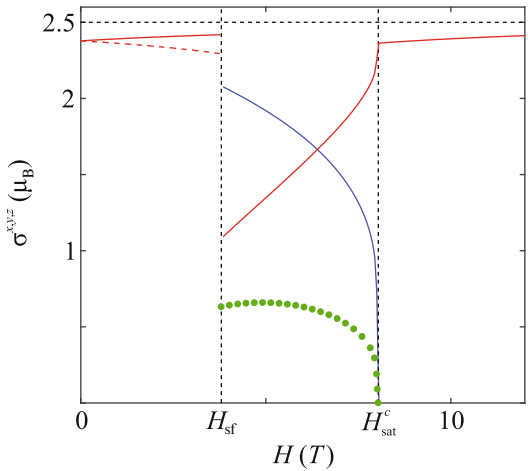

**Fig. 4.** (Color online) (Left panel) (Red solid line) $\sigma_F^z$ and (red dashed line) $|\sigma_G^z|$ values in the phase I. (Middle panel) (Red solid line) $\sigma_F^z = \sigma_G^z$, (blue solid line) $\sigma_F^x = -\sigma_G^x$, and (green dotted line) $|\sigma_F^y|$ ($\sigma_F^y = \sigma_G^y$) in the phase II. (Right panel) (Red solid line) $\sigma_F^z = \sigma_G^z$ in the phase III at the temperature $T = 2$ K and the parameters as in Fig. 2a.

**3.** Corrections to the magnetizations of the sublattices, which are due to zero-point quantum fluctuations arising because the Néel state does not correspond to the exact eigenfunction of the Hamiltonian of the system, have been analyzed. It has been shown in the analytical form that the corrections do not depend on the magnetic field and cannot explain the anomalies found in the experiment.

**4.** After solving the dispersion relation, the renormalized expressions for critical fields $H_{sf}$ and $H_{sat}^c$ are obtained from the condition that the positive definiteness of the excitation spectrum is lost in the magnetic field.

**5.** Taking into account these relations, magnetic field dependences of the magnetization at different temperatures are plotted by means of the numerical solution of the system of self-consistent equations. The obtained properties qualitatively correctly reflect the features of field dependences in $MnBi_2Te_4$.

**6.** The application of the results of the theory is not limited to the description of the magnetic subsystem of $MnBi_2Te_4$, but can also be used to interpret the properties of anisotropic quasi-two-dimensional magnets with a triangular lattice in layers. At the same time, the extension of the theoretical results to ultrathin films, which are relevant in view of the possible implementation of the quantum anomalous Hall effect in them, requires correction, in particular, because the single-ion anisotropy operator is uniform in our consideration, whereas the surface effects in thin films can induce the inhomogeneity of the crystal field.

## FUNDING

This work was supported by the Russian Science Foundation (project no. 23-22-10021, https://rscf.ru/en/project/23-22-10021) and by the Krasnoyarsk Regional Fund of Science.

## CONFLICT OF INTEREST

The authors of this work declare that they have no conflicts of interest.

*Translated by L. Mosina*